# Quantitative Correlation between One-Dimensional Periodically Oscillating Detonation Waves and Two-Dimensional Regular Cellular Detonation Waves

Yunfeng Liu[1,2]

1. Institute of Mechanics, Chinese Academy of Sciences, Beijing 100190, China
2. School of Engineering Science, University of Chinese Academy of Sciences, Beijing 100049, China

liuyunfeng@imech.ac.cn

**Abstract:** The objective of this study is to establish a quantitative relationship between one-dimensional (1D) periodically oscillating detonation waves and two-dimensional (2D) regular cellular detonation waves. Numerical simulations were conducted for stoichiometric hydrogen–air detonation waves using the conservative Euler equations coupled with a single-step overall chemical reaction model. The evolution of both 1D periodically oscillating detonation waves and 2D regular cellular detonation waves was analyzed through flux-vector analysis. The results reveal that the period of forming a 2D detonation cell is equivalent to the period of the 1D periodically oscillating detonation wave. Furthermore, 2D regular cellular detonation waves exhibit two distinct time scales: (1) the collision period of triple-points, and (2) the ignition delay time of the heated gas behind the incident shock wave. These two time scales reach a balance for regular cellular detonation waves. The findings demonstrate that the ignition delay time is the key physical parameter for establishing a direct quantitative correlation between the two systems.

## 1. Introduction

A detonation wave is a type of supersonic combustion wave. At present, two major theoretical frameworks exist: the Chapman–Jouguet (CJ) detonation theory and the Zeldovich–von Neumann–Döring (ZND) model [1,2]. Both theories describe detonation waves as possessing a one-dimensional, steady structure. However, in reality, detonation waves are inherently three-dimensional and unsteady. The detonation front is composed of an incident shock, a Mach stem, and transverse waves, with their intersections forming triple-points. The trajectories of these triple-points leave diamond-shaped cellular structures on the wall.

The detonation cell size is an important parameter for characterizing the chemical reaction kinetics of detonations [3]. Recently, Monnier et al. conducted experimental investigations on three-dimensional detonation cell structures [4]. Zhao et al. [5] and Sharp et al. [6] performed statistical analyses of detonation cell sizes. Numerous numerical simulations have been carried out to study the dynamics of detonation waves, exploring how parameters such as activation energy and grid resolution influence cell size [7–17]. These studies have revealed that activation energy affects the ignition delay time of chemical reaction models, and that detonation cell size exhibits a quantitative correlation with ignition delay time [18,19].

The instability of one-dimensional detonation waves is also a fundamental theoretical problem. Many researchers have applied shock-capturing schemes to numerically investigate the instability of 1D detonations [20–25]. Ng et al. [26,27] employed the Euler equations with a single-step overall reaction model to simulate the nonlinear dynamics of 1D detonations. Their studies showed that as activation energy increases, a detonation transitions from stable to unstable, then evolves into periodic detonations, non-periodic oscillatory detonations, and eventually chaotic states. Kasimov et al. [28] and Andrew et al. [29] used shock-fitting methods and high-order WENO schemes to study 1D detonation instabilities. Powers [30] provided a comprehensive review of 1D detonation instabilities.

To date, no research has successfully established a quantitative relationship between one-dimensional periodically oscillating detonations and two-dimensional regular cellular detonations [25]. Han et al. [31,32] made preliminary attempts to address this problem. Liu [33,34] first proposed the flux-vector analysis method for analyzing detonation dynamics. The objective of this study is to establish a quantitative connection between these two systems. First, the flux-vector analysis method is applied to examine the mechanism of 2D detonations, obtaining the period required to form a complete cell. Next, the mechanism of 1D periodically oscillating detonation is investigated to determine their oscillation period. It is found that these two periods are equal, thereby establishing a quantitative link between these two systems.

## 2. Governing Equations and Numerical Methods

The governing equations are based on the two-dimensional conservative Euler equations coupled with a single-step overall chemical reaction model. The effects of viscosity, molecular diffusion, and heat conduction are neglected.

$$\frac{\partial U}{\partial t} + \frac{\partial F}{\partial x} + \frac{\partial G}{\partial y} = S \tag{1}$$

$$U = \begin{pmatrix} \rho \\ \rho u \\ \rho v \\ \rho e \\ \rho Z \end{pmatrix} \quad F = \begin{pmatrix} \rho u \\ \rho u^2 + p \\ \rho uv \\ (\rho e + p)u \\ \rho u Z \end{pmatrix} \quad F = \begin{pmatrix} \rho v \\ \rho uv \\ \rho v^2 + p \\ (\rho e + p)v \\ \rho v Z \end{pmatrix} \quad S = \begin{pmatrix} 0 \\ 0 \\ 0 \\ 0 \\ \dot{\omega} \end{pmatrix} \tag{2}$$

$$e = \frac{p}{(\gamma - 1)\rho} + \frac{1}{2}(u^2 + v^2) + Zq \tag{3}$$

$$p = \rho RT \tag{4}$$

$$\dot{\omega} = -K\rho Z \exp\left(-\frac{CE_a(\gamma - 1)}{RT}\right) \tag{5}$$

$$\gamma(Z) = \frac{\gamma_1 R_1 Z/(\gamma_1 - 1) + \gamma_2 R_2 (1 - Z)/(\gamma_2 - 1)}{R_1 Z/(\gamma_1 - 1) + R_2 (1 - Z)/(\gamma_2 - 1)} \tag{6}$$

$$R(Z) = R_1 Z + R_2 (1 - Z) \tag{7}$$

In the governing equations, $\rho$, $p$, $T$, $u$, $v$, $e$, $q$, $R$, $\gamma$, $Z$ denote density, pressure, temperature, velocity, specific internal energy, specific heat release, gas constant, specific heat ratio, and the reaction progress variable, respectively. In the single-step chemical reaction model, $\dot{\omega}$ represents the chemical reaction mass production rate, $A$ is the pre-exponential factor, and $E_a$ is the activation energy. In the numerical study, the activation energy is varied by adjusting the parameter $C$ in Eq. (5). The detonation products are modeled with variable specific heat ratio and gas constant, allowing a more accurate representation of their thermodynamic states. In Eqs. (6) and (7), the subscripts 1 and 2 correspond to reactants and products, respectively.

The evolution of the conserved variables within each time step when an explicit time-marching method being used is described Eqs. (8)–(15). For a grid point, the pressure increment $\Delta p$ over a single time step depends on the algebraic sum of three contributions: the convective term, the kinetic energy term, and the heat-release term. A positive convective term corresponds to a compression wave, whereas a negative convective term indicates a rarefaction wave. The kinetic energy term is in phase with the convective term and arises from numerical interpolation; at the detonation front, this term is negative, signifying that part of the total energy is converted into kinetic energy. The heat-release term is positive; however, the exothermic chemical reaction in a single time step does not necessarily result in a pressure increment, which depends on whether the convective term corresponds to compression or rarefaction during that step.

$$U^{n+1} = U^n - \frac{\partial F^n}{\partial x}\Delta t - \frac{\partial G^n}{\partial y}\Delta t + S^n \Delta t \tag{8}$$

$$\Delta(\rho e) = (\rho e)^{n+1} - (\rho e)^n = -\frac{\partial(\rho e + p)u}{\partial x}\Delta t - \frac{\partial(\rho e + p)v}{\partial y}\Delta t \tag{9}$$

$$\rho e = \frac{p}{(\gamma - 1)} + \frac{1}{2}\rho(u^2 + v^2) + \rho Z q \tag{10}$$

$$\Delta p = p^{n+1} - p^n = \left(\Delta(\rho e) - \Delta\left(\frac{1}{2}\rho(u^2 + v^2)\right) - \Delta(\rho Z q)\right)(\gamma - 1) \tag{11}$$

$$\Gamma = \Delta(\rho e) = -\frac{\partial(\rho e + p)u}{\partial x}\Delta t - \frac{\partial(\rho e + p)v}{\partial y}\Delta t \tag{12}$$

$$\Delta Q = -\Delta(\rho q Z) = \rho^n Z^n q - \rho^{n+1} Z^{n+1} q \tag{13}$$

$$\Delta K = -\Delta\left(\frac{1}{2}\rho(u^2 + v^2)\right) = \frac{1}{2}\rho^n\left((u^n)^2 + (v^n)^2\right) - \frac{1}{2}\rho^{n+1}\left((u^{n+1})^2 + (v^{n+1})^2\right) \tag{14}$$

$$\Delta p = (\Gamma + \Delta K + \Delta Q)(\gamma - 1) \tag{15}$$

Numerical simulations of two-dimensional cellular detonation waves were carried out. To reduce computational cost, a moving computational domain was employed, such that the detonation front remained fixed at the right boundary of the domain. The aspect ratio of the computational domain was set greater than four. The domain was initially filled with a stoichiometric hydrogen–air mixture at 300 K and 0.1 MPa. Ignition was initiated by imposing high-temperature, high-pressure unreacted gas at the left boundary, and the detonation wave propagated from left to right. The activation energy was varied in different cases, while remaining constant within a given simulation.

The parameters of the single-step global chemical reaction model are as follows: $Z_1 = 1.0$, $Z_2 = 0$, $\gamma_1 = 1.4$, $\gamma_2 = 1.244$, $R_1 = 398.5 \text{J/(kg·K)}$, $R_2 = 348.9 \text{J/(kg·K)}$, $q = 3.5 \text{MJ/kg}$, $E_a = 4.794/(\gamma_2 - 1) \text{MJ/kg}$, $K = 7.5 \times 10^9 \text{s}^{-1}$. Details of the model can be found in Refs. [18,19]. The numerical scheme employs a second-order ENO method combined with a third-order TVD Runge-Kutta time integration scheme. The convective fluxes are evaluated using the Steger-Warming flux splitting method. In the two-dimensional simulations, a uniform grid with a resolution of 20 μm is applied throughout the computational domain, while in the one-dimensional simulations, a finer uniform grid of 5 μm is used. Mirror reflection boundary conditions are imposed at the upper and lower walls, and extrapolation boundary conditions are applied at the left and right boundaries.

## 3. Results and Discussion

### 3.1 Dynamics of Triple-Point Movement (C=1.0)

Taking an activation energy of C=1.0 as the baseline case, Fig. 1 illustrates the evolution process of the detonation cell. After ignition by the high-temperature, high-pressure region (yellow area on the left), an overdriven detonation wave is first formed at the early stage. The cells of the overdriven detonation are very small. After a period of evolution, the detonation wave reaches a stable state and develops into a regular cellular structure, with an average cell length of 12 mm and an average cell width of 8 mm. As shown in Fig. 1, the average period required to form a complete cell is 6 μs.

Specifically, the time from the collision of triple-points to their reflection at the wall is 3 μs, and the time from wall reflection to the next collision of these two triple-points is also 3 μs.

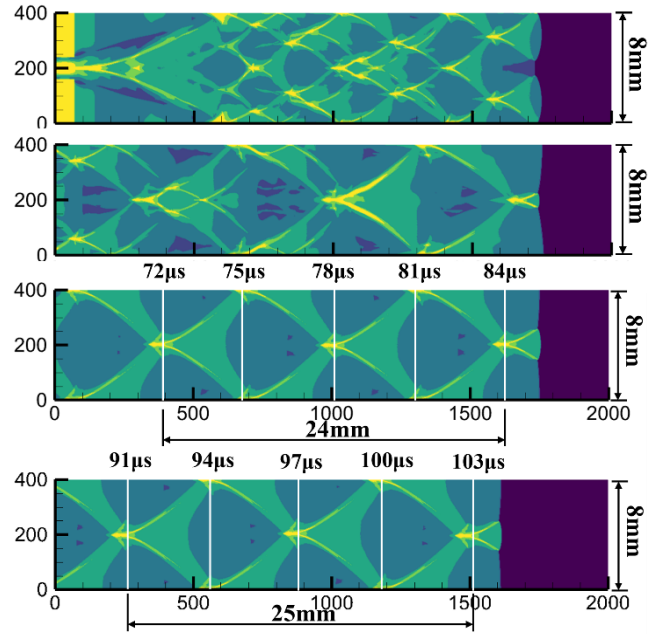

Figure 1. Evolution of the detonation cell structure (C=1.0)

Figure 2 presents the pressure and flux-vector contour plots of the detonation wave over a complete cycle from t = 78 μs to t = 84 μs. For clarity, the chemical heat release within a single time step is shown by black lines, shock waves by blue shading, and rarefaction waves by red shading. The maximum flux-vector magnitude shown is ±0.3 MJ/m³, and the maximum heat release shown is 0.2 MJ/m³.

At t = 78 μs, the two triple-points collide. The flux-vector contours show that, at the Mach stem, the chemical heat release is coupled with the shock (black lines overlapping with the blue shading), and no strong rarefaction is present (absence of red shading). In contrast, at the incident shock, the chemical heat release decouples from the shock and couples with the rarefaction wave (black lines overlapping with red shading). Since the rarefaction is stronger than the heat release, the net effect is a rarefaction wave.

At t = 79 μs, the two triple-points move toward the upper and lower walls, the Mach stem weakens, and the heat release has decoupled from the shock. However, at the local triple-point region (point A), although heat release and shock are decoupled (black lines and blue contours separated), no strong rarefaction is present (absence of red shading), and this region exhibits approximately isochoric combustion and generates pressure increasement. At t = 80 μs, the Mach stem continues to weaken. At the triple-point location (point A), a strong rarefaction wave emerges, the heat release couples with the rarefaction and the Mach stem degenerates into an incident shock. At t = 81 μs, the triple-points reflect from the wall and generate a new Mach stem. At t = 82 μs, the heat release decouples from the incident shock, though the rarefaction wave near the triple point is relatively weak. At t = 84 μs, the two triple-points collide again, forming a new Mach stem. At the Mach stem, the heat release couples once more with the shock. Thus, the detonation wave completes one cycle and forms a complete cell.

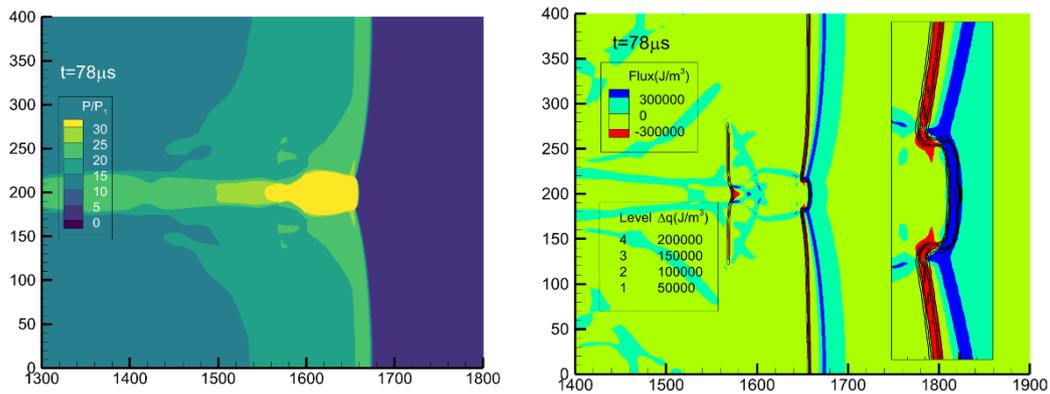

(a) t=78μs

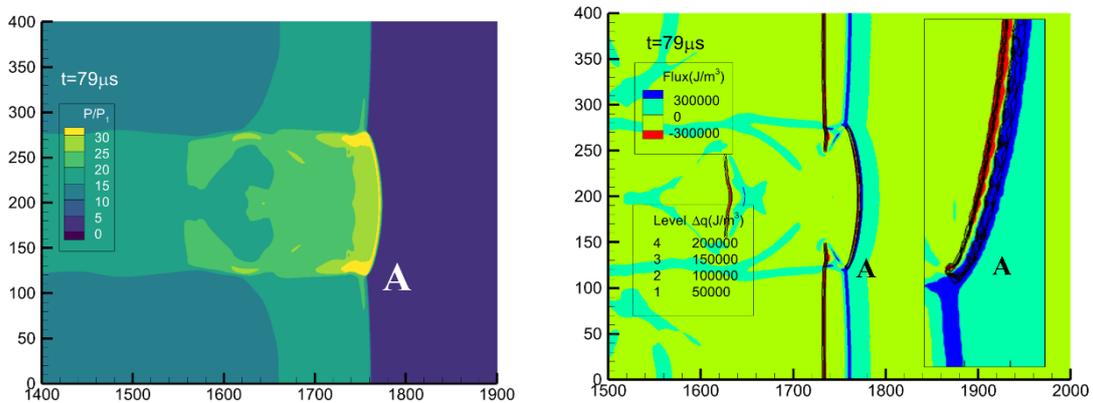

(b) t=79μs

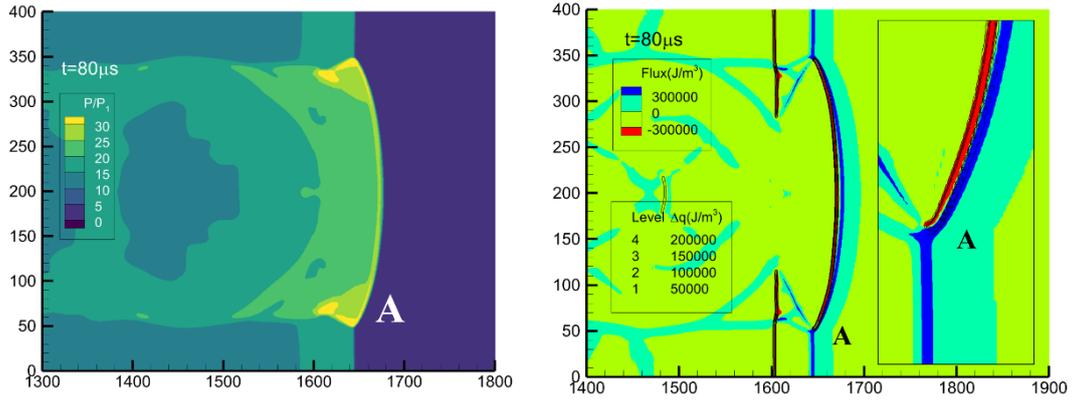

(c) t=80μs

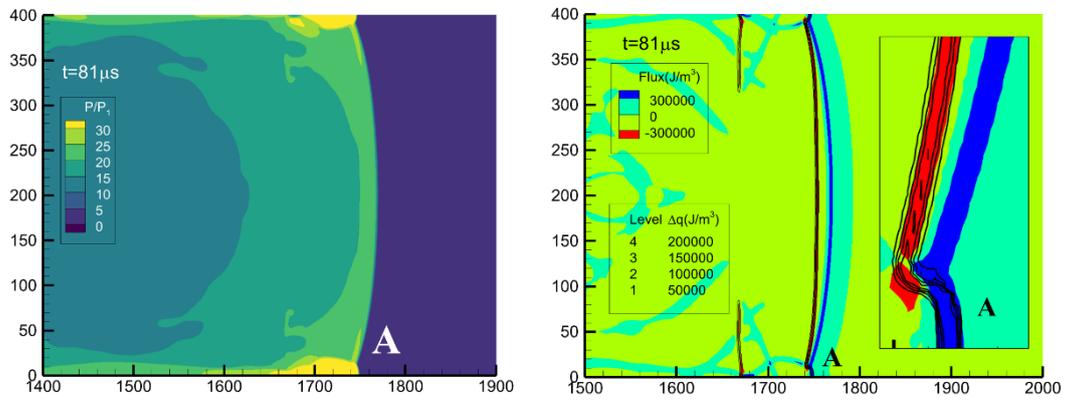

(d) t=81μs

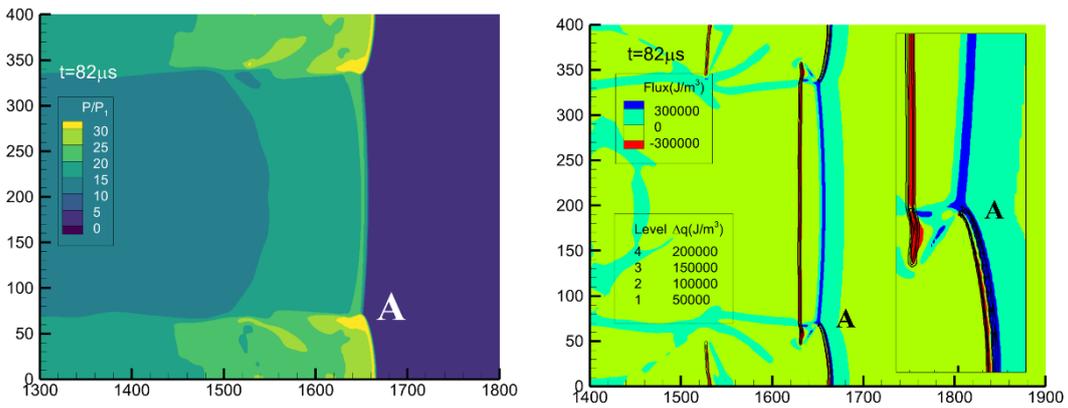

(e) t=82μs

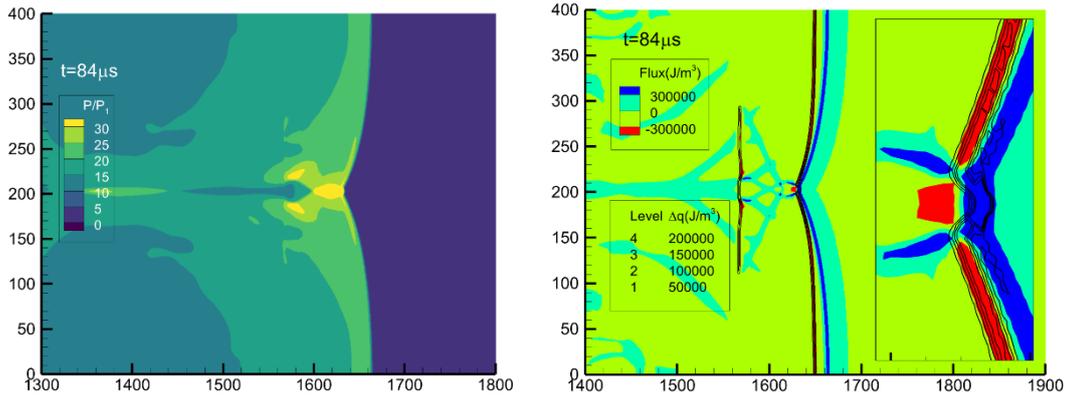

(f) t=84μs

Figure 2. Flux-vector and pressure contour plots (C=1.0)

Because the two triple-points differ in strength, the detonation cell gradually becomes asymmetric over time, with the stronger triple point slowly consuming the weaker one. Figure 3 illustrates the unsteady evolution process of the cell structure. As shown in Fig. 3, at t = 173 μs the two triple-points merge into a single triple-point. However, by t = 186 μs, a new triple-point is generated near the lower wall, and the flow field again contains two triple-points, forming a complete cell. As time progresses, the number of triple-points alternates between one and two.

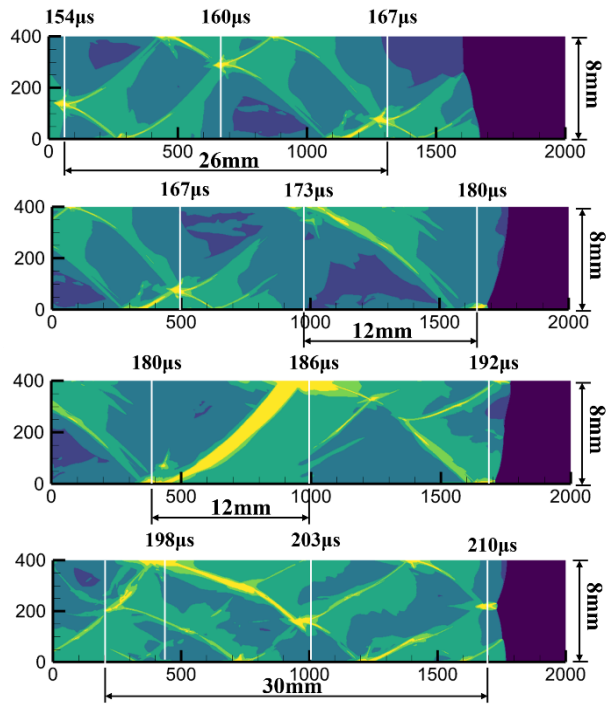

Figure 3. Evolution of the detonation cell structure (C=1.0)

Figure 4 presents the pressure and flux-vector contour plots corresponding to this unsteady new triple-point evolution process. As shown in Fig. 4, at t = 185 μs, the rarefaction wave near the incident shock weakens (narrowing of the red region at point A). Part of the chemical heat release couples with the shock (overlap of black lines and blue shading), producing a local pressure rise and forming a new triple-point. This new triple-point originates from the auto-ignition mechanism of the shocked mixture behind the incident shock. At t = 186 μs, this triple-point continues to strengthen, and the local rarefaction wave disappears (absence of red shading at point A), indicating an isochoric combustion process. At t = 187 μs, the two triple-points collide, and at t = 188 μs, the flow field again returns to a state with two triple-points.

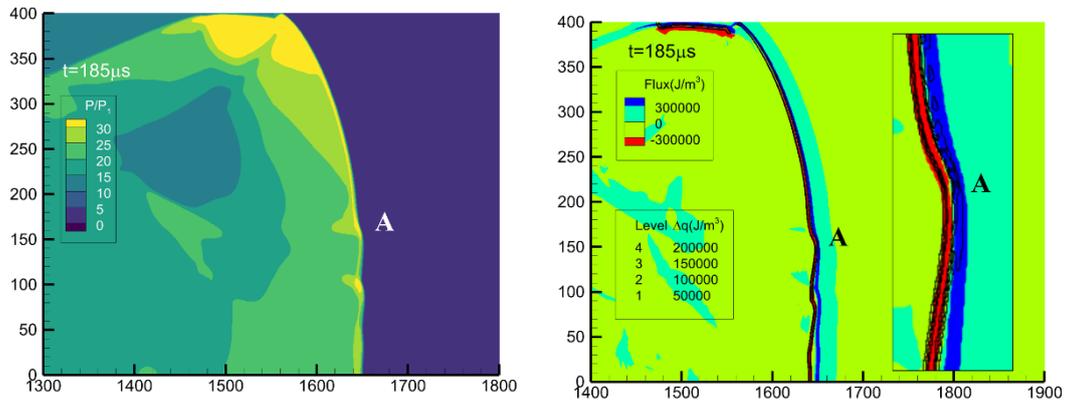

(a) t=185μs

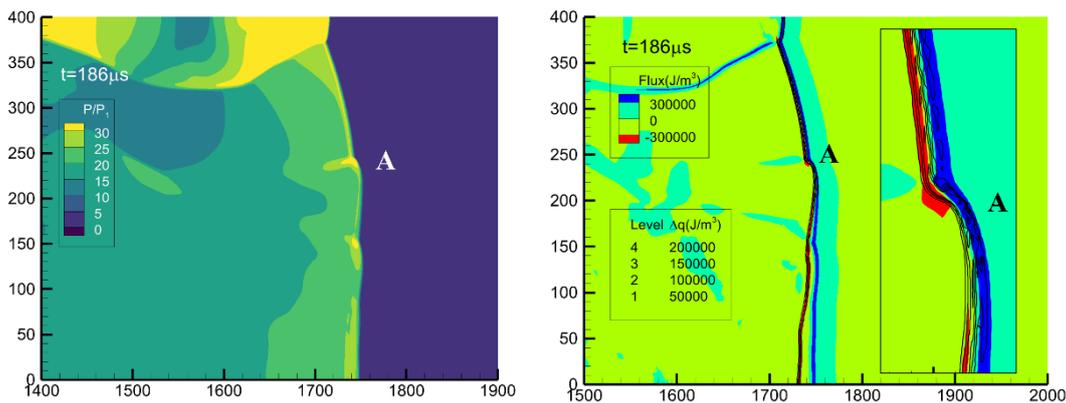

(b) t=186μs

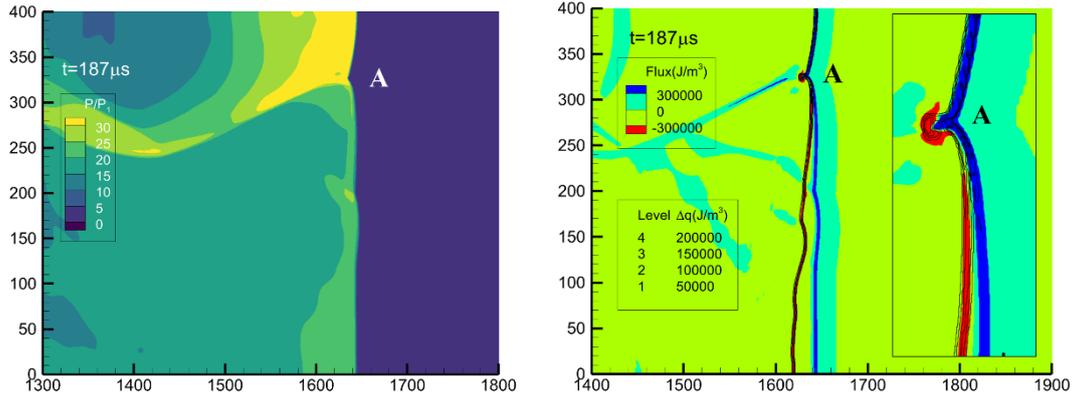

(c) t=187μs

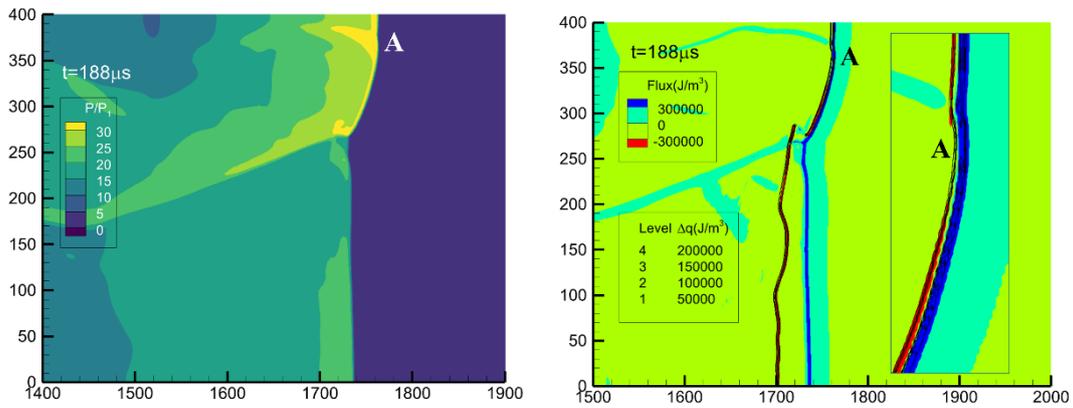

(d) t=188μs

Figure 4. Flux-vector and pressure contours (C=1.0)

During the propagation of detonation waves, the motion of triple-points is governed by two characteristic time scales: (1) the collision period of the triple-points, and (2) the ignition delay time of the shocked mixture behind the incident shock. As shown in Fig. 1, the time required for two triple-points to travel from reflection at the upper and lower walls (which can be regarded as collisions with additional triple-points) to their next collision is 3 μs. The single-step overall reaction model used in this study is pressure-independent, with an ignition delay time of 2.8 μs at 1500 K (the post-shock temperature corresponding to Mach 4.85). These two time scales are therefore equal. The propagation mechanism of 2D regular cellular detonations is the balance of these two time scales: the period of triple-point collision coincides with the auto-ignition time of the preheated mixture.

If these two time scales are not equal, two scenarios arise: either too many or too few triple-points exist. In the case of excessive triple-points, the stronger ones gradually consume the weaker ones, leading to a reduction in the number of triple-points. Conversely, if there are too few triple-points, once the preheated mixture behind the incident shock reaches its ignition delay time without a triple-point collision occurring, a new triple-point is generated through auto-ignition. As shown in Fig. 3, between t = 180 μs and t = 186 μs, the ignition delay time of the preheated gas near the lower wall is exceeded, triggering an auto-ignition event that produces a new triple-point.

**3.2 Relationship between Cell Size and Ignition Delay Time (C=0.7-1.0)**

The detonation cell size varies with activation energy, and a quantitative correlation exists between the cell size and the ignition delay time of the chemical reaction model used. Figure 5 shows the cell sizes obtained at different activation energies, while Figure 6 presents the corresponding ignition delay times from the single-step overall model. It is evident that as activation energy decreases, the ignition delay time shortens, and the detonation cell size becomes smaller. Table 1 summarizes the statistical results of cell size and ignition delay time at different activation energies. As seen in Table 1, the two are directly proportional, consistent with the conclusions of Refs. [18,19]. Table 1 also indicates that for C = 1.0, the numerically simulated cell size is in good agreement with experimental measurements.

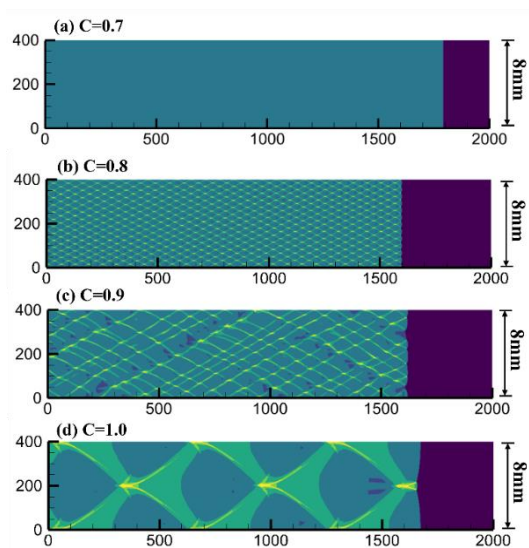

Figure 5. Detonation cell sizes at different activation energies (C=0.7-1.0)

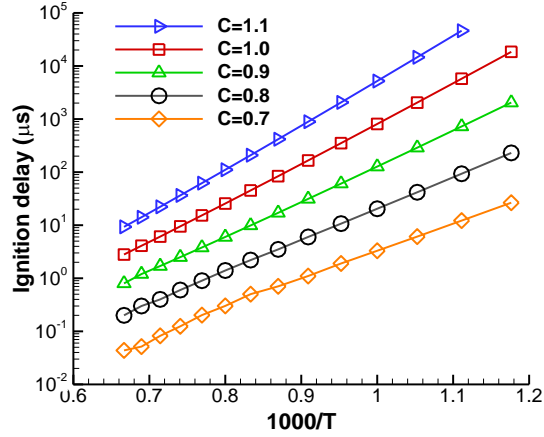

Figure 6. Ignition delay times at different activation energies (C=0.7-1.1)

**Table I. The ignition delay time and detonation cell size**

| Active energy | Ignition delay time at 1500K (μs) | Average cell width (mm) |
| --- | --- | --- |
| C=0.7 | 0.04 | No cell |
| C=0.8 | 0.2 | 0.62 |
| C=0.9 | 0.8 | 1.33 |
| C=1.0 | 2.8 | 8 |
| C=1.1 | 9.3 | Failure |
| Experimental cell width (mm) | 8.18[35]; 8[36]; 15.1[37]; 11.4[38]; 9.2[39]; 7[40]; 5.05[41] | |

Figure 7 shows the pressure and flux-vector contours at lower activation energies. For the case of C = 0.7, because the activation energy is very low, the chemical heat release couples with the shock front across the entire detonation wave. No rarefaction waves are observed, the pressure front remains flat, and the wave exhibits the characteristics of a one-dimensional steady detonation, with no cellular structure. For the case C = 0.8, the heat release decouples from the shock, but the rarefaction wave at the Mach stem is very weak. The detonation cells are therefore small and highly regular. For the case of C = 0.9, the cell size becomes larger and less regular. However, due to the relatively lower activation energy, a greater number of triple-points exist, and at their locations the rarefaction waves remain weak. These detonations are also called weakly unstable detonations, being different from periodically oscillating detonation waves.

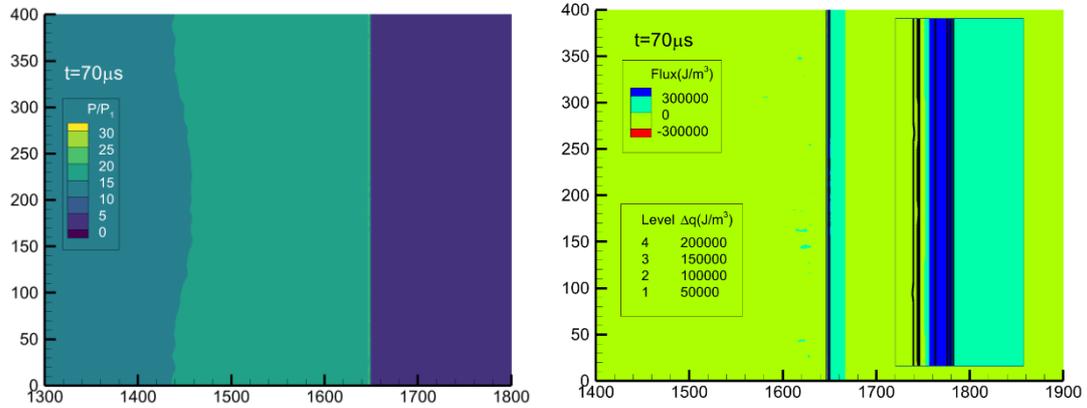

(a) C=0.7

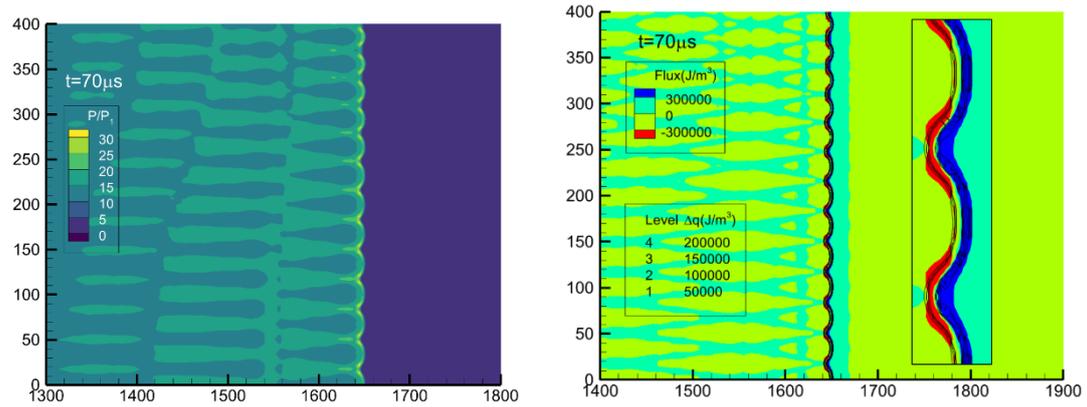

(b) C=0.8

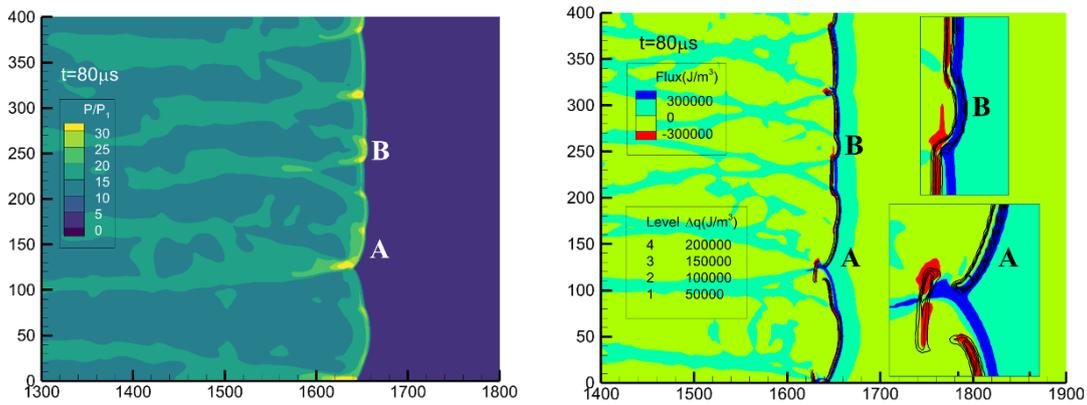

(c) C=0.9

Figure 7. Pressure and flux-vector contours at low activation energies (C=0.7-0.9)

## 3.3 Mechanism of Detonation Wave Quenching (C=1.1)

When the activation energy is increased to C = 1.1, the detonation wave can no longer sustain self-propagation. Figure 8 shows the cell structure during the initiation and extinction process. After ignition, the detonation wave propagates approximately one cell length before extinguishing. Figure 9 presents the pressure and flux-vector contours at t = 10 μs. At this moment, the two triple-points collide; however, due to the high activation energy, no chemical heat release occurs at the collision site (absence of black lines). At other locations, the chemical heat release couples only with rarefaction waves, and thus cannot generate a pressure rise. This demonstrates that detonation self-sustainable propagation is maintained through the collisions of triple-points. If such collisions fail to trigger chemical reactions, the detonation will be extinguished.

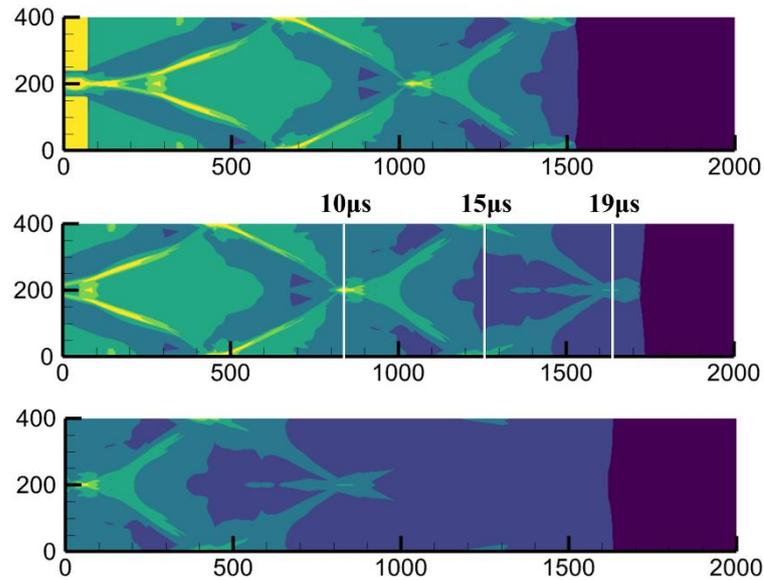

Figure 8. Cell structure during the detonation extinction process (C=1.1)

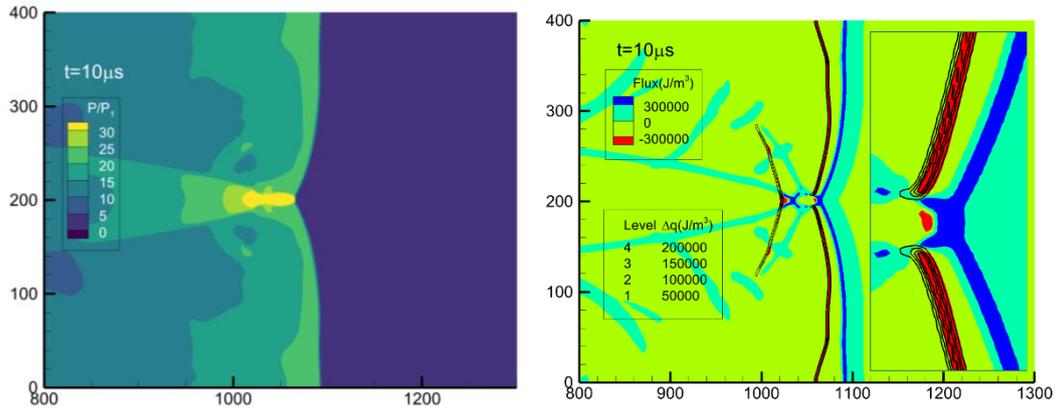

Figure 9. Pressure and flux-vector contours (C=1.1)

### 3.4 Quantitative Relationship between 1D Periodic Detonation Oscillations and 2D Regular Cellular Detonations (C=1.0)

Using the same single-step overall reaction model and computational framework, one-dimensional periodically oscillating detonation waves were also simulated. Figure 10 shows the temporal variations of the pressure ratio peak and velocity for the 1D periodically oscillating detonation wave, together with a comparison to the pressure peak and velocity along the centerline of the 2D regular cellular detonation wave. As illustrated in Fig. 10, these two periods are quantitatively identical: one oscillation period of the 1D detonation wave corresponds exactly to one cell period of the 2D detonation wave.

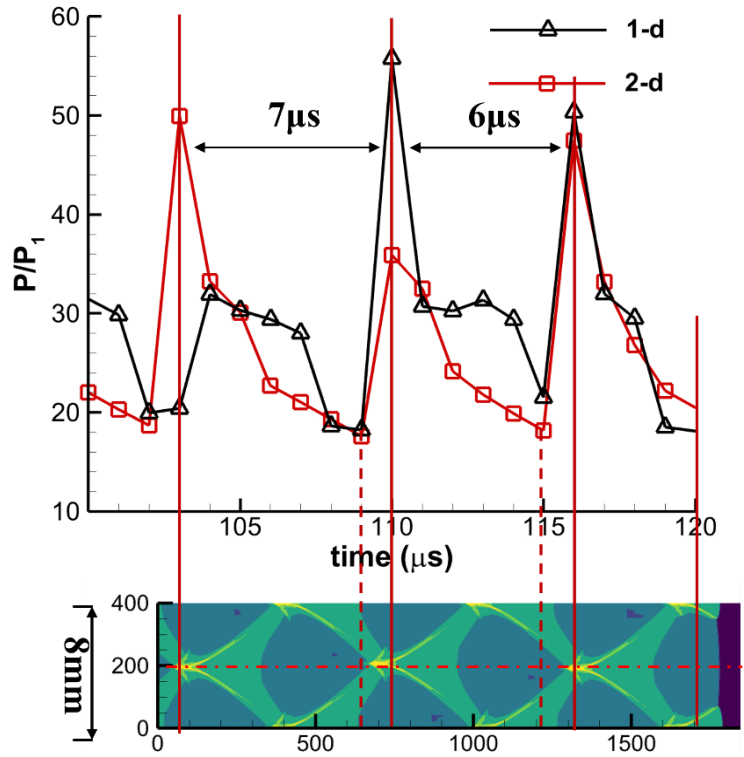

(a) Peak Pressure Ratio

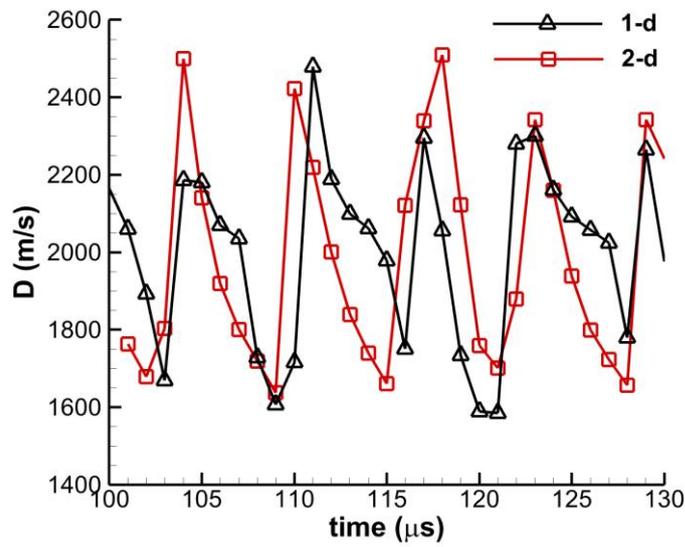

(b) Detonation Wave Velocity

Figure 10. Comparison of 1D periodic detonation oscillations and 2D regular cellular detonations

Figure 11 illustrates the evolution of flux-vector components for 1D periodically oscillating detonation wave over one cycle from t = 110 μs to t = 116 μs. In the figure, the symbols Γ, Q, and K denote the peak points of the convective term, heat release term, and kinetic-energy term, respectively.

At t = 110 μs, the shock and heat release peak coincide, marking the beginning of the cycle. At t = 111 μs, the detonation wave weakens, with the heat release peak lagging the shock by 0.01 mm, though partial overlap still exists. At t = 112 μs, the separation distance increases to 0.015 mm. At t = 113 μs, the detonation approaches a critical decoupling state. At t = 114 μs, decoupling begins and the rarefaction wave associated with the heat release front strengthens. At t = 115 μs, the detonation is fully decoupled, with the separation distance increasing to 0.06 mm. The heat release completely overlaps with the rarefaction wave, resulting in isobaric combustion. At t = 116 μs, auto-ignition occurs in the preheated mixture behind the leading shock wave. Finally, at t = 116.1 μs, the heat release recouples with the leading shock wave gain, the pressure reaches its peak, and a new oscillation cycle begins.

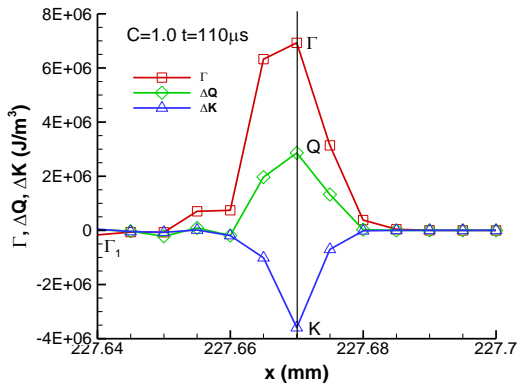
(a) t=110μs

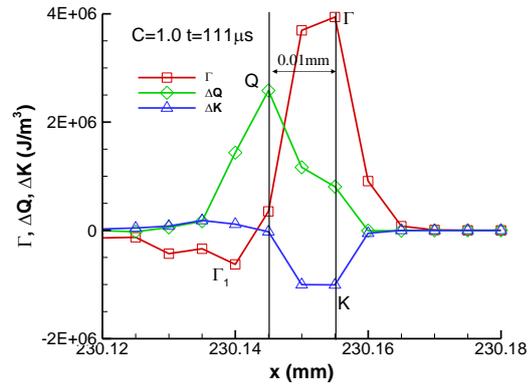
(b) t=111μs

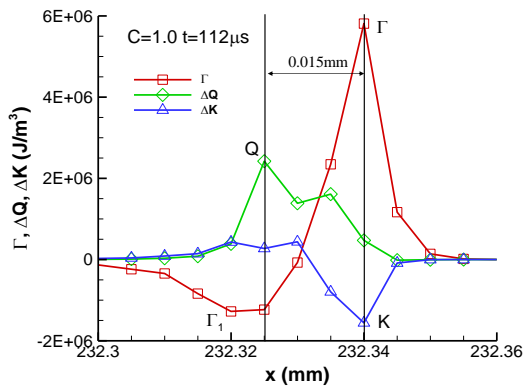
(c) t=112μs

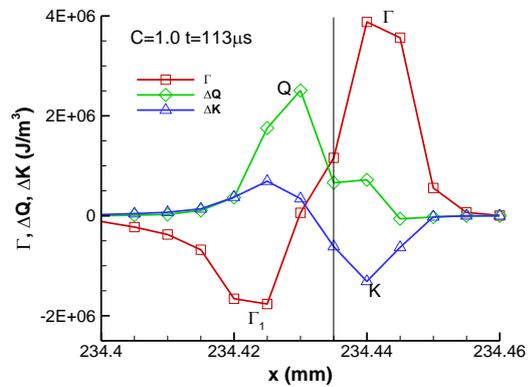
(d) t=113μs

(c) t=112μs  (d) t=113μs

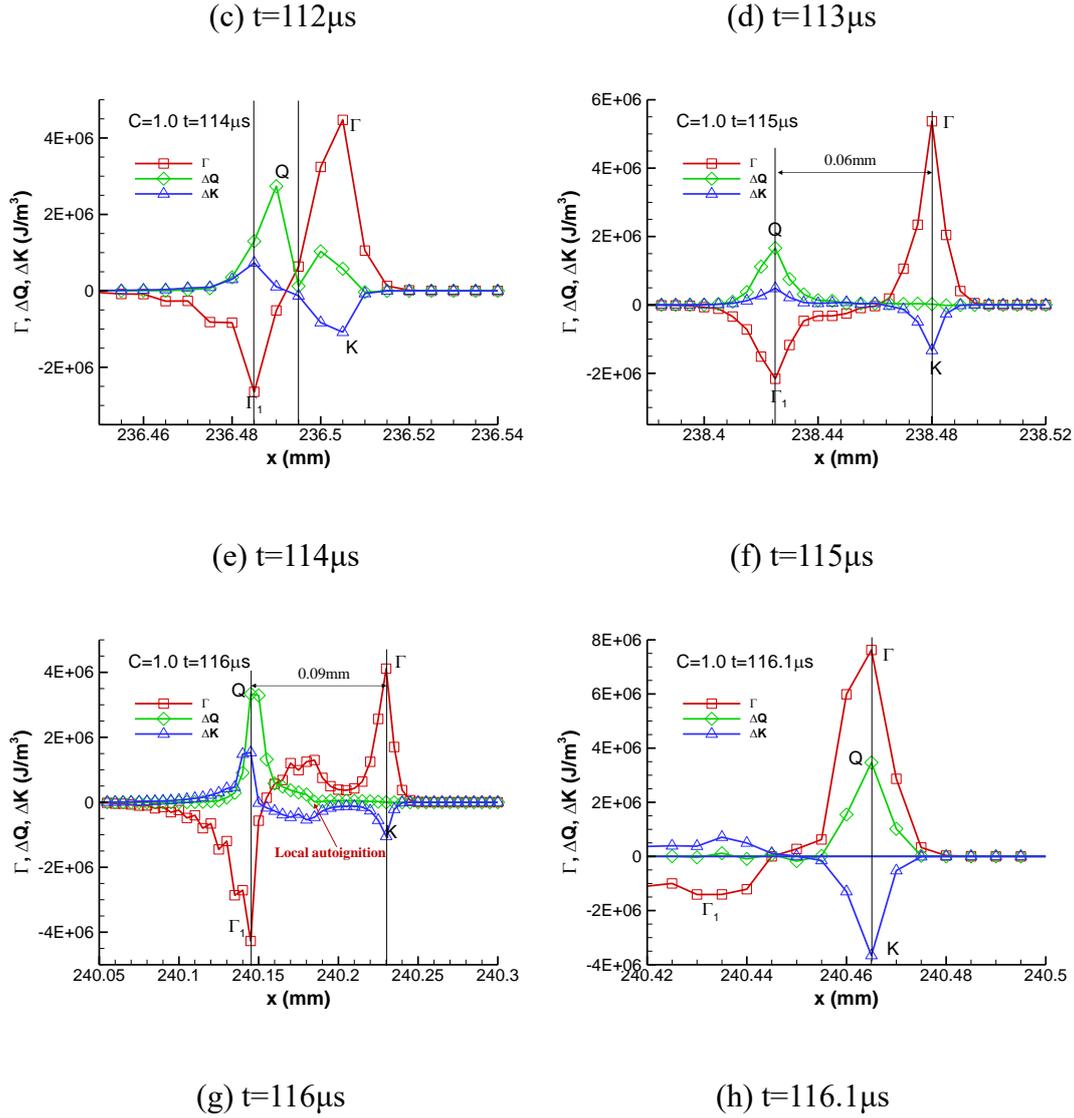

(e) t=114μs  (f) t=115μs

(g) t=116μs  (h) t=116.1μs

Figure 11. Flux-vector evolution of 1D periodically oscillating detonation (C=1.0)

As shown in Fig. 11, the average period of the one-dimensional periodically oscillating detonation wave is 6 μs. The time from the critical decoupling state at t = 113 μs to the onset of auto-ignition at t = 116 μs is 3 μs, which is exactly equal to the ignition delay time of the single-step overall reaction model at 1500 K (the post-shock temperature corresponding to Mach 4.85). Similarly, the period of one cell in the two-dimensional regular cellular detonation wave is also 6 μs, indicating that the two are quantitatively identical.

In summary, the propagation mechanism of the 1D periodically oscillating detonation wave can be summarized as: auto-ignition ➔ overdriven detonation ➔ decoupling ➔ auto-ignition. In contrast, the mechanism of the 2D regular cellular detonation wave is: triple-point collision ➜ Mach stem formation ➜ Mach stem attenuation into an incident shock➜triple-point collision. During the propagation of the 2D regular cellular detonation, the motion period of the triple-points is equal to the ignition delay time of the shocked mixture behind the incident shock. Thus, ignition delay time serves as the key physical parameter that directly links the 1D periodically oscillating detonation wave to the 2D regular cellular detonation wave.

## 4. Conclusion

Using the flux-vector analysis method, the propagation mechanisms of two-dimensional regular cellular detonations and one-dimensional periodically oscillating detonations in stoichiometric hydrogen–air mixtures were investigated. The results reveal that two characteristic time scales govern the 2D regular cellular detonation: (1) the collision period of the triple points and (2) the ignition delay time of the preheated gas behind the incident shock. These two time scales are balanced. The period required to form a complete cell in the 2D detonation is equal to the oscillation period of the 1D periodically oscillating detonation, both corresponding to the ignition delay time of the chemical reaction model. These two systems are quantitatively linked through the ignition delay time, which serves as the key physical parameter connecting their dynamics.